\documentclass[12pt]{article}
\usepackage{cite,amssymb}

\begin{document}

\title{\bf Anisotropic Inflation in a 5D Standing Wave Braneworld and Dimensional Reduction}

\author{{\bf Merab Gogberashvili,$^{1,2}$} \ {\bf Alfredo Herrera--Aguilar,$^{3,4}$} \ {\bf Dagoberto} \\
{\bf Malag\'on--Morej\'on$^3$ \ and \ Refugio Rigel Mora--Luna$^3$}\\
$^1$Andronikashvili Institute of Physics,\\
6 Tamarashvili St., Tbilisi 0177, Georgia\\
$^2$Javakhishvili State University, \\
3 Chavchavadze Ave., Tbilisi 0128, Georgia\\
$^3$Instituto de F\'{\i}sica y Matem\'{a}ticas,\\
Universidad Michoacana de San Nicol\'as de Hidalgo \\
Edificio C--3, Ciudad Universitaria, CP 58040, Morelia, Michoac\'{a}n, M\'{e}xico\\
$^4$Centro de Estudios en F\'{\i}sica y Matem\'aticas B\'asicas y Aplicadas, \\
Universidad Aut\'onoma de Chiapas,
\\Calle 4a Oriente Norte 1428, Tuxtla Guti\'errez, Chiapas, M\'exico\\ \\
{\sl E-mails: gogber@gmail.com, alfredo.herrera.aguilar@gmail.com,}\\
{\sl malagon@ifm.umich.mx \ and \ rigel@ifm.umich.mx} }

\maketitle

\begin{abstract}
We investigate a cosmological solution within the framework of a 5D standing wave braneworld model generated by gravity coupled to a massless scalar phantom-like field. By obtaining a full exact solution of the model we found a novel dynamical mechanism in which the anisotropic nature of the primordial metric gives rise to i) inflation along certain spatial dimensions, and ii) deflation and a shrinking reduction of the number of spatial dimensions along other directions. This dynamical mechanism can be relevant for dimensional reduction in string and other higher dimensional theories in the attempt of getting a 4D isotropic expanding space-time.

\end{abstract}

\vskip 0.5cm

In order to study the cosmology of a braneworld model one must first completely solve both the bulk and the brane field equations. Only when such a complete solution is at hand the brane dynamics can be consistently analyzed using the brane projection formalism \cite{SMS}. In general, this is a difficult tusk and we know very few cosmological braneworld models that possess an exact solution, especially in the case of anisotropic metrics \cite{bwcosmolanisot}. As a consequence, there is a considerable amount of research in the literature that obtained several results by making some assumptions concerning the behavior of the Weyl term on the brane when the full solution to the bulk metric was unknown \cite{assumps}. The main mathematical difficulty in anisotropic case consists in obtaining generalizations of the $AdS_5$ space-time, which incorporate anisotropy on the cosmological 3-brane and are necessarily non-conformally flat. In this framework, it is interesting to construct new anisotropic braneworld models that supplement or generalize the existing ones, and attempt to describe more realistic cosmological braneworlds and/or to explore other aspects of higher-dimensional gravity that have been not considered until now.

Recently it was introduced the model where at some `critical time' in a 10-dimensional universe, three out of nine spatial directions start to expand, leading to a space with $SO(3)$ symmetry instead of $SO(9)$ \cite{KEK}. The result was obtained within the framework of a matrix non-perturbative formulation of type IIB superstring theory by numerical simulations on a supercomputer. We want to show that a similar dynamical mechanism can be formulated within a 5D standing wave braneworld model generated by gravity coupled to a phantom-like scalar field. Namely, by starting with an anisotropic 5D metric {\it ansatz} and leaving it evolve for large times, certain spatial dimensions of the 3-brane will shrink to zero-size while others will expand in an accelerated way for certain values of the parameters of the solution.


In this article we shall look for a cosmological solution within a generalization of the 5D standing wave braneworld model reported in \cite{Wave,CQG,GMM} which introduces the scale factors $F_1(t)$ and $F_2(t)$ in the metric {\it ansatz} as follows:
\begin{equation} \label{metric}
ds^2 = e^{2a|r|}\left[ dt^2 - F_1^2(t) e^{u(t,r)} \left(dx^2 + dy^2 \right) - F_2^2(t)e^{-2u(t,r)}dz^2\right] - dr^2~,
\end{equation}
where the constant $a$ corresponds to the brane width towards the extra dimension $r$ and the function $u(t,r)$ represents gravitational waves in the bulk. The peculiarity of the model (\ref{metric}) is that the brane, located at $r = 0$, possesses anisotropic oscillations and sends the waves $u(t,r)$ into the bulk (as in \cite{GMS}) and three spatial dimensions can evolve
(expanding or contracting) in time through the scale factors $F_1(t)$ and $F_2(t)$.

The 5D Einstein equations of the model with the real, massless, phantom-like scalar field $\phi$, the 5D cosmological constant $\Lambda$, and the brane as a source can be written in the form:
\begin{equation} \label{Einstein}
R_{MN} =  \frac{2}{3} g_{MN} \Lambda + 8\pi G \left[-\partial_M \phi \partial_N \phi + \widetilde{T}_{MN} + \frac{2}{3} g_{MN} \sigma(\phi,t)\delta(r) \right] ~,
\end{equation}
where capital Latin indexes refer to 5D space-time, $G$ is the 5D Newton constant, $\sigma(\phi,t)$ is a brane tension which, in general, depends on the scalar field and time, and
\begin{equation} \label{tideT}
\widetilde{T}_{MN} = T_{MN} - \frac{1}{3}g_{MN}T^A_A
\end{equation}
is the 5D energy-momentum tensor of the brane. In \cite{CQG} it was shown that, in the absence of the scale factors $F_1(t)$ and $F_2(t)$, the anisotropic nature of the metric, together with the presence of terms proportional to $\delta(r)$ in the Ricci tensor, necessarily requires the presence of an anisotropic energy-momentum tensor of a brane `effective fluid', which constitutes a mixture of a vacuum fluid (characterized by the brane tension) and an anisotropic matter fluid with different stresses along different directions. Thus, we take the energy-momentum tensor of the brane as:
\begin{eqnarray} \label{tensormixto}
T^{M}_{N}=\mbox{\rm diag}[\lambda_0,\lambda_1,\lambda_2, \lambda_3,0]\delta{(r)}~, ~~~~~ \lambda_1 = \lambda_2~,
\end{eqnarray}
where $\lambda_0(t,r)$ denotes the energy density of the 'effective fluid' and $\lambda_m(t,r)$ ($m=1,2,3$) correspond to the anisotropic stresses along the spatial directions of the brane. Thus, the non-zero components of the tensor (\ref{tideT}) are:
\begin{eqnarray}
\widetilde{T}^t_t &=& \frac{2\lambda_0-2\lambda_1-\lambda_3}{3}~\delta(r)~, \nonumber \\
\widetilde{T}^x_x &=&\widetilde{T}^y_y = \frac{-\lambda_0+\lambda_1-\lambda_3}{3}~\delta(r)~, \nonumber\\
\widetilde{T}^z_z &=& \frac{-\lambda_0-2\lambda_1+2\lambda_3}{3}~\delta(r)~,\\
\widetilde{T}^r_{r} &=& - \frac{\lambda_0+2\lambda_1+\lambda_3}{3}~\delta(r)~. \nonumber
\end{eqnarray}

By computing the non-zero components of the Ricci tensor for the metric (\ref{metric}), the Einstein equations (\ref{Einstein}) read:
\begin{eqnarray} \label{Ricci}
R^t_t &=& - e^{-2a|r|}\left[\frac {3}{2}\dot u ^2 + 2\frac{\ddot F_1}{F_1}+ \frac{\ddot F_2}{F_2} + 2\dot u\left(\frac{\dot F_1}{F_1}-\frac{\dot F_2}{F_2}\right)\right]+ 4a^2 + 2a\delta(r) = \nonumber \\
&=& \frac{2}{3}\Lambda + 8\pi G\left[- e^{-2a|r|}\dot\phi\,^2 + \widetilde{T}^t_t + \frac{2}{3}\sigma\,\delta(r)\right]~, \nonumber \\
R^x_x&=&R^y_y = - e^{- 2a|r|}\left[ \frac {1}{2}\ddot u + \frac{\ddot F_1}{F_1} + \frac{1}{2}\dot u\left( 2\frac{\dot F_1}{F_1}+\frac{\dot F_2}{F_2}\right) + \left(\frac{\dot F_1}{F_1}\right)^2 + \frac{\dot F_1}{F_1}\frac{\dot F_2}{F_2}\right] +\nonumber \\
&+& 4a^2 + 2a~{\rm sgn}(r)u' + 2a\delta(r)+ \frac {1}{2} u'' =
\frac{2}{3}\Lambda + 8\pi G\left[\widetilde{T}^x_x + \frac{2}{3}\sigma\,\delta(r)\right]~, \nonumber \\
R^z_z &=& - e^{-2a|r|}\left[ -\ddot u +\frac{\ddot F_2}{F_2} - \dot u\left(2\frac{\dot F_1}{F_1}+\frac{\dot F_2}{F_2}\right) + 2\frac{\dot F_1}{F_1}\frac{\dot F_2}{F_2}\right] +\\
&+&  4a^2 - 4a~{\rm sgn}(r)u' + 2a\delta(r) - u'' =
\frac{2}{3}\Lambda + 8\pi G\left[\widetilde{T}^z_z + \frac{2}{3}\sigma\,\delta(r)\right]~, \nonumber \\
R^r_{r} &=& \frac 32 u'^2 + 4a^2 + 8a\delta(r) =
\frac{2}{3}\Lambda + 8\pi G \left[-\phi'\,^2 + \widetilde{T}^r_r + \frac{2}{3}\sigma\,\delta(r)\right]~, \nonumber\\
R^r_{t} &=& u'\left(\frac{3}{2}\dot u + \frac{\dot F_1}{F_1}-\frac{\dot F_2}{F_2}\right) = 8\pi
G\dot\phi\phi' ~, \nonumber
\end{eqnarray}
where ${\rm sgn}(r)$ is the sign function, overdots and primes mean derivatives with respect to $t$ and $r$, respectively.

Due to the last two expressions of the system (\ref{Ricci}), it follows that:
\begin{eqnarray}\label{rr-rt}
\Lambda = 6 a^2~, \nonumber \\
\sqrt{\frac{16\pi G }{3}}~\phi'(t,r) = u'(t,r)~, \\
\sqrt{\frac{16\pi G }{3}}~\dot\phi (t,r)=\dot u(t,r) + \frac{2}{3}\left(\frac{\dot F_1}{F_1}-\frac{\dot F_2}{F_2}\right)~. \nonumber
\end{eqnarray}
The last relation of this expressions generalizes a previous result obtained in \cite{Wave,CQG} where a simple proportionality between $u$ and $\phi$ was found in the absence of $F_1(t)$ and $F_2(t)$.

The Klein-Gordon equation for the phantom-like scalar field under (\ref{metric}) reads:
\begin{equation}\label{phi}
\ddot\phi + \left(2\frac{\dot{F_1}}{F_1}-\frac{\dot{F_2}}{F_2}\right)\dot\phi - e^{2a|r|}\left[\phi'' + 4a~{\rm sgn}(r) \phi'\right] =
e^{2a|r|}\frac{\partial\sigma}{\partial\phi}~\delta(r)~.
\end{equation}
By taking into account this equation in the $xx$- and $zz$-components of the Einstein equations (\ref{Ricci}) and combining the resulting expressions with the $tt$-component (upon the last relation of (\ref{rr-rt})) we obtain a system of differential equations for the scale factors
$F_1(t)$ and $F_2(t)$:
\begin{eqnarray}
\frac{\ddot{F_2}}{F_2}+2\frac{\dot{F_1}}{F_1}\frac{\dot{F_2}}{F_2}= 0~, \nonumber\\
\frac{\ddot{F_1}}{F_1}+\left(\frac{\dot{F_1}}{F_1}\right)^2 + \frac{\dot{F_1}}{F_1}\frac{\dot{F_2}}{F_2}= 0~, \label{F1F2}\\
2\frac{\ddot{F_1}}{F_1}+\frac{\ddot{F_2}}{F_2} -\frac 23 \left(\frac{\dot{F_1}}{F_1} - \frac{\dot{F_2}}{F_2}\right)^2 = 0~. \nonumber
\end{eqnarray}
This system has the single exact solution:
\begin{equation}\label{expsoln}
F_1(t) \sim e^{Ht}~, ~~~~~ F_2(t) \sim e^{-2Ht}~,
\end{equation}
where $H$ is a constant.

After substituting (\ref{expsoln}) into the equation for the scalar field (\ref{phi}) we get:
\begin{equation}\label{phi2}
\ddot\phi - e^{2a|r|}\left[\phi'' + 4a~{\rm sgn}(r) \phi'\right] = e^{2a|r|}\frac{\partial\sigma}{\partial\phi}~\delta(r)~,
\end{equation}
which precisely coincides with the corresponding equation of \cite{Wave,CQG} when $r\ne 0$. By inserting the expressions for $F_1(t)$ and $F_2(t)$ from (\ref{expsoln}) into the last expression
of (\ref{rr-rt}) we obtain:
\begin{eqnarray}\label{dotu+2H}
\sqrt{\frac{16\pi G }{3}}~\dot\phi (t,r)=\dot u(t,r) + 2H~,
\end{eqnarray}
which necessarily implies that the second time derivative of $u$ is proportional to that of $\phi$:
\begin{eqnarray}\label{ddotu}
\sqrt{\frac{16\pi G }{3}}~\ddot\phi (t,r) = \ddot u(t,r)~.
\end{eqnarray}
This fact, together with the relations (\ref{rr-rt}), shows that the relevant differential equation for $u(t,r)$:
\begin{equation}\label{eqnu}
\ddot u - e^{2a|r|}\left[u'' + 4a~{\rm sgn} (r) u'\right] = \sqrt{\frac{16\pi G }{3}} ~ e^{2a|r|} \frac{\partial\sigma}{\partial\phi} ~ \delta(r)~,
\end{equation}
has exactly the same form as the equation for $\phi (t,r)$ (\ref{phi2}) when $r\ne 0$ and possesses the standing wave solution \cite{Wave,CQG}:
\begin{eqnarray}
\label{gsolu} u(t,r) = \sin(\omega t) e^{-2a|r|}\left[{\cal A}~J_2\left( \frac{\omega}{|a|} e^{-a|r|} \right) + {\cal B}~Y_2\left( \frac{\omega}{|a|} e^{-a|r|} \right)\right],
\end{eqnarray}
where ${\cal A},{\cal B}$ and $\omega$ represent the amplitudes and frequency of the wave, and $J_2$ and $Y_2$ are $2^{nd}$ order Bessel functions of first and second kind, respectively. Moreover, by taking into account the associated Sturm-Liouville problem (which keep a normalizable function along the fifth dimension) for the equation (\ref{eqnu}) when $r\ne 0$, we must set ${\cal B}=0$ and we are left with the following solution \cite{CQG}:
\begin{equation} \label{solnu}
u(t,r) = {\cal A}~ \sin (\omega t) e^{-2a|r|} J_2\left( \frac{\omega}{|a|} e^{-a|r|} \right)~,
\end{equation}
where the parameter $a$ should be positive definite $a>0$. Furthermore, by inserting (\ref{solnu}) into the relation (\ref{dotu+2H}) and integrating it with respect to time we get the following expression for the phantom-like scalar field:
\begin{equation} \label{solnphi}
\phi(t,r) = \sqrt{\frac{3}{16\pi G}}~ \left[ {\cal A}~ \sin (\omega t) e^{-2a|r|} J_2\left(\frac{\omega}{|a|} e^{-a|r|} \right) + 2Ht \right]~,
\end{equation}
up to an arbitrary additive constant. Note that the last term was absent in the previous result obtained in \cite{Wave,CQG}, where the scale factor functions $F_1(t)$ and $F_2(t)$ were trivial ($H=0$). We see that the amplitude of the ghost-like field $\phi$ increases/decreases linearly with time depending on the sign of the constant $H$.

It is worth noticing that the brane at $r = 0$ is at rest in one of the nodes of the bulk standing wave described by the oscillatory function $u(t,r)$. This is accomplished by the fine tuning:
\begin{equation}\label{X}
\frac {\omega}{a} = X ~,
\end{equation}
where $X$ is one of the zeros of the Bessel function $J_2$ in (\ref{solnu}). When the frequency of the standing waves $\omega$ is much larger than the frequencies associated with the energies of particles on the brane we can perform a time averaging of the oscillating exponents of the metric (\ref{metric}), which enter the equation for matter fields. In \cite{GMM} it was explicitly demonstrated that the resulting $r$-dependent functions form potential wells and can provide pure gravitational localization of all kind of matter fields on then brane.

Once we have solve the Einstein - Klein-Gordon system in the bulk we can proceed to solve the corresponding equations on the brane.

The junction condition on the brane for the Klein-Gordon equation implies that the jump of the first derivative of the scalar field $\phi$ at $r=0$, denoted by $[\phi']$, reads
\begin{equation} \label{junctionKG}
\left[\phi'\right] = - \left.\frac{\partial\sigma}{\partial\phi}~\right|_{r=0}~.
\end{equation}
On the other hand, as pointed out in \cite{CQG}, the junction conditions on the brane coming from
the Einstein equations establish a relation between the jump $[u']$, the parameter $a$ and the
anisotropic stresses $\lambda_m$. These restrictions are consistently satisfied by the following
expressions for the energy density $\lambda_0$, the stresses $\lambda_m$, the jump $[u']$ and the
brane tension $\sigma(\phi,t)$:
\begin{eqnarray} \label{tensions}
\lambda_0&=&-\frac{3a}{4\pi G}~, \nonumber \\
\lambda_1&=& \lambda_2 ~ = -\frac{{\cal A}\omega\,\sin(\omega t)} {8 \pi G}J_1|_X - \frac{3a}{4 \pi G}~, \nonumber \\
\lambda_3&=& \frac{{\cal A}\omega\,\sin(\omega t)}{4 \pi G} J_1|_X - \frac{3a}{4 \pi G}~, \label{match1} \\
\,[u']&=& \frac{16 \pi G}{3}(\lambda_1-\lambda_3)=-2{\cal A}\omega \,\sin(\omega t) J_1|_X =
-\sqrt{\frac{16\pi G }{3}}~\left.\frac{\partial\sigma}{\partial\phi}\,\right|_{r=0}~, \nonumber \\
\sigma|_{r=0}&=& 0~, \nonumber
\end{eqnarray}
where $J_1$ is a $1^{st}$ order Bessel function of first kind. We see that the energy density of the `effective fluid' $\lambda_0$ is constant. The spatial components of the brane tension $\lambda_m$ can decompose as follows:
\begin{equation} \label{lambda}
\lambda_m = \lambda_0 + \pi_m~, ~~~~~ (m=1,2,3)
\end{equation}
and can be physically interpreted according to a covariant analysis developed from the viewpoint of a brane-bound observer \cite{Maartens,KG}: the non-constant part of (\ref{lambda}) represents the anisotropic stresses that oscillate along the spatial directions $x^m$:
\begin{equation} \label{anisotropic_t}
\pi_1 = \pi_2 = -\frac 12 \pi_3 = -\frac{{\cal A}\omega\,\sin(\omega t)} {8 \pi G}J_1|_X ~.  \\
\end{equation}
Although the contribution to the brane tension given by the scalar
field is zero
\begin{equation}
\sigma(\phi|_{r=0},t) = 0~,
\end{equation}
it influences the space-time on the brane through the junction
conditions (\ref{junctionKG}) and (\ref{tensions}), since the jump
of the derivative of the scalar field $[\phi']$, and hence the
jump of the derivative of the function $[u']$, is not zero.

An expression for $\sigma(\phi,t)$ that satisfies the junction condition (\ref{junctionKG}) and
the last equality of (\ref{tensions}) together with the fine tuning condition (\ref{X}), which
amounts to set to zero the phantom scalar field $\phi$ on the zeroes of the Bessel function $J_2$,
is given by:
\begin{equation} \label{sigma}
\sigma(\phi,t) = \sum_{n=1}^{\infty} C_n\,\left[\phi-\phi(0)\right]^n\,\sin(\omega t)~,
\end{equation}
where $C_n$ for ($n\ge 2$) are arbitrary constants and $C_1$ satisfies the following relation
\begin{equation} \label{Cn}
C_1 = \sqrt{\frac{3}{4\pi G}}{\cal A}\,\omega\,J_1|_X~
\end{equation}
which is a sort of fine tuning among the constants $C_1$, ${\cal A}$, $\omega$ and the value of
the Bessel function $J_1$ at the zeroes of the Bessel function $J_2$.

Finally, the scale factor functions $F_1(t)$ and $F_2(t)$ given by (\ref{expsoln}), the normalizable metric function $u(t,r)$ (\ref{solnu}), the scalar field $\phi (t,r)$ (\ref{solnphi}), the expressions for the anisotropic source tensions together with the jump $\,[u']$ on the brane given by relations (\ref{tensions}) and the brane tension $\sigma$ (\ref{sigma}), conform the full solution for the coupled system of Einstein and ghost-like scalar field equations, leading to the background metric:
\begin{equation} \label{metric1}
ds^2 = e^{2a|r|}\left[ dt^2 - e^{[u(t,r) + Ht]} \left(dx^2 + dy^2 \right) - e^{ - 2 [u(t,r) + Ht]}dz^2\right] - dr^2~.
\end{equation}
We notice that since $u(t,r)$ oscillates in time, for large time intervals, the properties of the space-time (\ref{metric1}) crucially depend on the sign of the constant $H$:

$1).$ For the positive constant,
\begin{equation}
H > 0~,
\end{equation}
the space-time (\ref{metric1}) expands exponentially in the $x$ and $y$ directions and squeezes in the $z$ direction. This means that in a macroscopical time interval the 3-brane surface at $r=0$ will shrink into a 2-brane. Therefore, the 3-brane will effectively have two space-like dimensions. At the same time, the amplitude of the ghost-like field (\ref{solnphi}) increases with time.

$2).$ In the case
\begin{equation}
H < 0~,
\end{equation}
in the metric (\ref{metric1}) $z$-distances will expand and the $(x-y)$-plane will shrink, leading to a 1-string, and in this case we shall have just one spatial dimension in the 3-brane. Simultaneously, the amplitude of the ghost-like field (\ref{solnphi}) will decrease in time.

This mechanism of dynamical asymmetric dimensional reduction of multi-dimensional surfaces could be useful for string models when obtaining a 4D isotropic expanding space-time from a higher-dimensional anisotropic universe.


To summarize, in this paper we have investigated an anisotropic cosmological solution arising
within a 5D standing wave braneworld supported by a ghost-like bulk scalar field, which is also
allowed to evolve in time. The consistent solution of the bulk and brane field equations leads to
a dynamical mechanism under which an initial anisotropic metric evolves in time and yields a
universe expanding along certain spatial dimensions, and contracting and shrinking to zero size
along other spatial directions. In principle, this novel mechanism provides a dynamical arena in
which a higher-dimensional world renders a 4D expanding isotropic universe, while other spatial
dimensions effectively disappear after a suitable time interval. The cosmological consequences for
theories that involve extra dimensions like string theory and the braneworld paradigm are
remarkable.

\medskip


\noindent {\bf Acknowledgments:} The research of MG was supported by the grant of Shota Rustaveli
National Science Foundation $ST~09.798.4-100$. This research was supported by grants CIC-UMSNH and
CONACYT $60060-J$. RRML and DMM acknowledge PhD grants from CONACYT and UMSNH, respectively. AHA
thanks SNI for support.


\end{document}